\documentclass[twocolumn,secnumarabic,amssymb, nobibnotes, prb, supebrscriptaddress]{revtex4-2}

\usepackage{xcolor}
\usepackage{amsmath}
\usepackage{physics}
\usepackage{graphicx}
\usepackage{wrapfig}
\usepackage{float}
\usepackage{subcaption}
\usepackage{booktabs}
\usepackage{multirow}
\usepackage{array}
\usepackage{setspace}
\graphicspath{{Figs/}}
\usepackage[colorlinks = true, linkcolor = blue, urlcolor  = blue, citecolor = blue, anchorcolor = blue]{hyperref}
\begin{document}

\title{Spin-orbit coupling tuned crossover of gaped and gapless topological phases in the chalcopyrite HgSnX$_2$ (X=N/P): An \textit{ab-initio} investigation}
\author {Surasree Sadhukhan}
\email{Corresponding author : surasree183212004@iitgoa.ac.in, surasree.sadhukhan@gmail.com}
\author{Sudipta Kanungo}
\email{sudipta@iitgoa.ac.in}
\affiliation{School of Physical Sciences, Indian Institute of Technology Goa, Goa-403401, India}
\date{\today}
\begin{abstract}

The coupling between electron orbital momentum and spin momentum, known as spin-orbit 
coupling (SOC), is a fundamental origin of a multitude of fascinating physical 
phenomena, especially it holds paramount significance in the realm of topological 
materials. In our work, we have predicted the topological phase in Hg-based 
chalcopyrite compounds using the first principles density functional theory. The 
initial focus was on HgSnN$_2$, revealing it to be a nonmagnetic Weyl semimetal, 
while HgSnP$_2$ displayed characteristics of a strong topological insulator. What 
makes our work truly unique is that despite both compounds having the same SOC 
strength, \textcolor{black}{arises from Hg}, they exhibit distinct topological phases 
due to the \textcolor{black}{distinct hybridization effect of the Hg-$5d$ and X-$p$ 
bands}. This finding can address \textcolor{black}{a significant factor, i.e., the effect of the band hybridization in deriving distinct topological phases, keeping the symmetry aspect intact.} Our results indicate that due to the presence 
of band hybridization between the dominant \textcolor{black}{X-$np$ orbitals n=2 and 3 for X=N and P respectively} and a minor contribution from \textcolor{black}{Hg-${5d}$}, we can tune the topological phase by manipulating SOC strength, \textcolor{black}{which equivalently achievable by chemical substitutions.} This investigation stands as a remarkable illustration of the unique roles that hybridization plays in sculpting the topological properties of these compounds while simultaneously preserving their underlying symmetries.

\end{abstract}
\maketitle
\section{Introduction}
The coupling between spin and orbital angular momentum in the relativistic framework, known as spin-orbit coupling (SOC), is closely intertwined with the emergence of novel phases in quantum materials\citep{krasovskii2015spin, fan2014quantifying, soumyanarayanan2016emergent,moon2013non}. This fascinating domain of SOC has broadened to encompass a plethora of phenomenon like Kitaev spin liquid systems\citep{huang2017interplay, catuneanu2018path, kitagawa2018spin}, unconventional magnetism in heavy elements\citep{kim2015kitaev}, superconductivity\citep{shick2019spin,chakrabortty2023effect}, Rashba splitting\citep{manchon2015new,caviglia2010tunable,governale2002quantum}, chiral magnetic interaction\citep{yu2023chirality,kim2013chirality, grytsiuk2020topological, banerjee2014enhanced}, spin-orbit torque (SOT)\citep{kim2017spin, brataas2014spin, zhu2019spin, baek2018spin, ciccarelli2016room, tesavrova2013experimental,lee2021efficient,song2021spin,li2019manipulation}, spin Hall effect\citep{kane2005quantum, hoffmann2013spin}, spin-orbit qubits\citep{nadj2010spin}, topological phases of matter\citep{hasan2010colloquium, qi2011topological, beugeling2012topological, zhang2022topological, kim2022three, young2011theoretical, island2019spin, kheirkhah2020first, fang2022ferromagnetic, rademaker2022spin, Dejean2022,tian2020spin} and many more. The importance of SOC in relation to topological phases is thoroughly recognized and meticulously documented, especially the incorporation of SOC introduces a band gap and leads to band inversion\citep{pesin2012spintronics,ezawa2015monolayer,bian2016topological, singh2018spin, howlader2020strong} and even SOC notably plays a significant role in instigating the breakage of band degeneracy\citep{tian2020spin}. The strength of SOC is an atomic property that is directly proportional to the fourth power of atomic number; nonetheless, some possible approaches, such as doping Pb$_{1-x}$Sn$_{x}$Se\citep{dziawa2012topological} and designing heterostructures\citep{mahfouzi2012spin,naimer2023twist}, have been highly appreciated to engineer the effects of SOC and utilize in topological-based spintronics devices\citep{soumyanarayanan2016emergent,belayadi2023spin}. Furthermore, both the magnitude and sign of the SOC strength can be manipulated as experimentally demonstrated in thallium nitride (TlN)\citep{sheng2014topological}. However, the precise and definitive connection between topological phases and SOC at a material-specific level remains a subject yet to be fully explored. Especially in the context of inter-metallic compounds, the hybridization between bands originating from distinct chemical elements can effectively account for the influence of SOC. In this scenario, the identification of the topological phase is not exclusively dictated by the particular element demonstrating high SOC. Instead, the compelling impact of hybridization with other elements, even those with lower SOC strength, comes into play, influencing the determination of the topological signatures.

To understand the direct impact of SOC in driving the topological phases, we have selected two sister compounds, HgSnN$_2$ and HgSnP$_2$, having the same crystalline symmetry, valence electrons, and heavy element (Hg), which accounts for the dominance of SOC strength.
The rationale behind such a choice of compounds is to ensure everything that determines the topology of the materials is the same except for the choice of the anion(N/P). Here it is crucial for meaningful comparisons in our investigation. Moreover, the study of SOC effects in topological phases of Hg-based heavy elements has found promising platforms in several cases, including HgTe quantum well, which has similarity with chalcopyrite\citep{zholudev2012magnetospectroscopy}, MTl$_4$Te$_3$ (M = Cd, Hg)\citep{li2022electronic}, Jacutingaite-family [M$_2$NX$_3$ (M =
Ni, Pt, and Pd; N = Zn, Cd, and Hg; and X = S, Se, and Te)]\citep{de2020jacutingaite}, two dimensional M$_3$C$_2$ (M = Zn, Cd, and Hg)\citep{liu2017two}, Hg-based materials HgS, HgSe, and HgTe\citep{virot2013engineering} etc., from both theoretical and experimental standpoints. 
Additionally, they provide insights that can aid in predicting potential Hg-based topological materials in our case. Moreover, it is also crucial to consider the experimental feasibility of compounds for a sensible choice of quantum material. The sibling compounds based on N and P in the chalcopyrite class have already been synthesized successfully, including ZnSiP$_2$\citep{mughal1969preparation}, ZnGeP$_2$\citep{verozubova2010growth, masumoto1966preparation}, ZnSnP$_2$\citep{rubenstein1968preparation}, CdSiP$_2$\citep{zawilski2010growth, zhang2012growth}, CdGeP$_2$\citep{vohl1979synthesis}, CdSnP$_2$\citep{buehler1971concerning,shirakata1990growth}, MgSiP$_2$\citep{trykozko1975ternary}, ZnSnN$_2$\citep{khan2020review}, II-Sn-N$_2$ (II= Ca, Mg, Zn)\citep{kawamura2021synthesis},  MgSnN$_2$\citep{greenaway2020combinatorial}, MgZrN$_2$\citep{rom2021bulk}, MnSiN$_2$\cite{esmaeilzadeh2006crystal} and many more. While our specific compound hasn't been experimentally synthesized yet, there is a significant likelihood of achieving experimental synthesis for these compounds in the future. It is worth highlighting the added advantages of small gap semiconductors with broken inversion symmetry have the potential to serve as promising candidates for various applications, including photovoltaic cells, optoelectronics (such as light-emitting diodes and lasers), thermoelectricity, and non-linear optical properties\citep{fan2014energetic,ohmer1998emergence,sadhukhan2020first}. Additionally, the ability to engineer a wide range of band gaps offers opportunities for developing band gap tuning devices driven by SOC.

In our investigation, we have followed the \textit{ab-initio} first principles method to identify two previously unexplored topological materials, namely HgSnN$_2$ and HgSnP$_2$. To the best of our knowledge, no previous investigations have been conducted to examine the topological properties of these compounds. Our findings reveal that HgSnN$_2$ identified as a nonmagnetic Weyl semimetal, showcasing essential topological features such as band inversion, Fermi arc, chirality, etc., in the presence of SOC. Similarly, the HgSnP$_2$ shows the characteristics of a strong topological insulator with a Z$_2$ number of [1,000] and an odd number of Dirac cones in its surface spectral function when SOC is incorporated. The distinctive aspect of our research lies in the fascinating observation that despite the atomic SOC strength being identical for both compounds, they showcase distinctly different topological phases: one is gapless, and another is gapped. Such disparity of nontrivial phase arises from the band hybridization of $N$/$P$ elements with $Hg$, resulting in their unique behavior of gapless or gaped band structure. To explore the tangible impact of SOC on hybridized bands, we have artificially modulated the strength of SOC in the original Hamiltonian. Intriguingly, our investigations revealed that by escalating the strength of SOC in HgSnN$_2$, we could induce a gap in the Weyl crossing, leading to a topological phase similar to that of the parent compound HgSnP$_2$. In a parallel manner, when we decrease the SOC strength in HgSnP$_2$, we observed the band gap closing, resulting in a Weyl crossing akin to the parent HgSnN$_2$ compound. This comparative analysis sheds light on the distinct roles played by hybridization in shaping their topological properties. \textcolor{black}{The importance of our present work lies in the fact that, as per the usual notion, the crossover from gap less to gaped topological phases can be achieved by tuning the governing symmetries; however, in the present case, we demonstrate how tuning SOC and hybridization can lead to the gap less to gap crossover in the non-trivial topological phases.}


\section{Crystal structure}

\begin{figure*}
\includegraphics[scale=0.4]{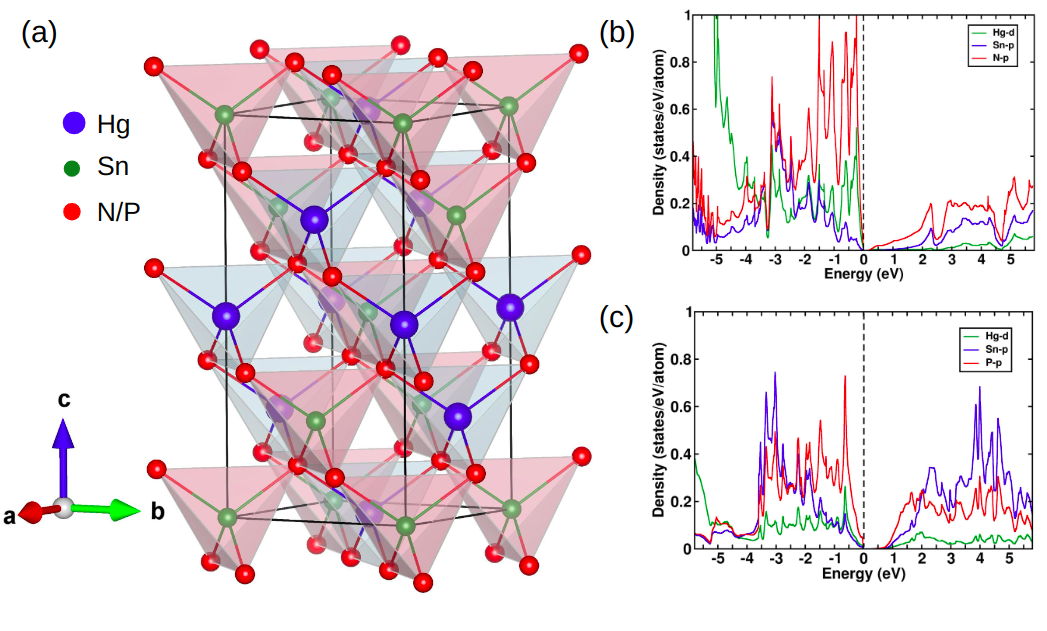}

\caption{(Color online)(a) The crystal structure of HgSnX$_2$ [X=N, P] is depicted, where Hg, Sn, and X atoms are represented by blue, green, and red spheres, respectively. \textcolor{black}{We utilized the VESTA package with a ball-and-stick representation to visualize the three-dimensional crystal structure\cite{momma2008vesta}.}(b) and (c) display the calculated orbital projected density of states (DOS) of HgSnX$_2$, where X=N and P, respectively. The green, blue, and red lines represent the partial DOS for \textcolor{black}{Hg-${5d}$, Sn-${5p}$, and X-${np}$ (n=2 and 3 for X=N and P respectively),} respectively. The energy scale is referenced to the Fermi energy (E$_f$), which is set at zero.}
\label{Fig-1-Struc-DOS}
\end{figure*}

\begin{table}
\begin{center}
\begin{tabular}{|c|c|c|c|c|}
\hline
 Material & \textit{a} ($\AA$) &  \textit{c} ($\AA$)& Ref. \\
\hline
    HgSnN$_2$       & 5.19 & 9.90  & our calculations\\
                    & 6.46 & - &  \citep{suh2004combinatorial}\\
                    
                    & 4.96 & 9.65   &\citep{basalaev2020simulation}\\

 \hline
         HgSnP$_2$    & 5.98 & 11.77 &  our calculations \\
                      & 6.17 & - &  \citep{suh2004combinatorial}\\
                      & 5.87 & 11.47 & \citep{basalaev2020simulation}\\
                      & 5.95 & 11.83  &  \citep{zhong2019first}\\
                      & 5.91 & -  &  \citep{jaffe1984theory}\\
\hline                   
\end{tabular}
\caption{A showcasing of the optimized lattice parameters \textit{a} and \textit{c} for both HgSnN$_2$ and HgSnP$_2$. The data provided includes our calculations and the corresponding values found in the literature, with appropriate references.}
\end{center}
\label{HGN}
\end{table}  

\begin{table*}
\centering
\begin{tabular}{|c|c|c|c|c|c|}
 \hline
   HgSnN$_2$ & Component of & HgSnP$_2$ &  Component of\\
    (0.016 THz) & displacement vector($\AA$) & (0.010 THz) & displacement vector ($\AA$)\\
\hline
Hg            & ( -0.11,  0.15, 0.00)& Hg& ( -0.01,   -0.01    0.18 )\\
\hline
     Sn         &  ( -0.09,  0.12, 0.00) & Sn & (-0.01,-0.00, 0.14)\\
\hline
 N          &  (-0.03,  0.04,  0.00 )&P & (-0.00,   -0.00,    0.07)\\


\hline                   
\end{tabular}
\caption{\textcolor{black}{The component of displacement vectors corresponding to Hg, Sn, and X (where X = N or P) are presented in the acoustic mode with the lowest frequency at the $\Gamma$ point.}}
\label{Acoustic}
\end{table*}

\begin{table*}
\centering
\begin{tabular}{|c|c|c|c|c|c|}
\hline
   HgSnN$_2$ & Component of &HgSnP$_2$ &  Component of\\
    (18.97 THz) & displacement vector($\AA$) & (9.71 THz) & displacement vector ($\AA$)\\
\hline
Hg            & (0.00, 0.00, 0.00)& Hg& (0.01, 0.01, 0.00) \\
\hline
     Sn         & ( 0.00, 0.00,  0.00)   & Sn & (-0.07, -0.05, -0.00)\\
\hline
 N          & ( 0.15, 0.00, -0.09)        &P & (0.12,   -0.05   -0.11) 
\\
           &  (0.05,   0.15,    0.09)       & & (-0.01,   0.13,    0.11)\\
    &      (-0.05,   -0.15,    0.09)   & & ( -0.01    0.13   -0.11)  \\
        &    ( -0.15,  -0.00   -0.09) & & ( 0.12,   -0.05    0.12) 
\\

 \hline

\end{tabular}
\caption{\textcolor{black}{The component of displacement vectors corresponding to Hg, Sn, and X (where X = N or P) are presented in the optical mode with the highest frequency at the $\Gamma$ point.}}
\label{Optical}
\end{table*}


HgSnX$_2$ (X=N, P) is a non-centrosymmetric intermetallic alloy belonging to the chalcopyrite class with space group I$\bar{4}$2d (space group no. 122). The unit cell possesses the body-centered tetragonal structure having a tetrahedral unit with the cation Hg, Sn, and the anion X (X=N, P) as depicted in Fig. \ref{Fig-1-Struc-DOS}(a). The optimized lattice parameters are listed in Table I. The Wyckoff positions of the three atoms are assigned the position as Hg atom at (0, 0, 0), Sn atom at (0, 0, 0.5), and X atom at ($x$, 0.25, 0.125), where $x$ referred as the anion displacement parameter. For HgSnN$_2$ and HgSnP$_2$, the values of $x$ are 0.7921 and 0.7548, respectively. We have observed slight deviations in the lattice parameters of both compounds in the literature compared to our optimized values, which we listed in Table I. Therefore, we have examined the lattice parameters versus energy plot very carefully and confirmed that the lattice parameters listed in Table I are energetically lower compared to any other choice of lattice parameters reported so far, as demonstrated in Fig. S1(a) and Fig. S1(b) for HgSnN$_2$. Similar analyses have been conducted for HgSnP$_2$, as shown in Fig. S1(d) and S1(e). \textcolor{black}{The calculated formation energies for HgSnN$_2$ and HgSnP$_2$ are -1.6374 eV/f.u. and -6.91705 eV/f.u., respectively, which show that the HgSnP$_2$ is more stable than HgSnN$_2$. } Since this compound has yet to be synthesized experimentally, the choice of lattice parameters is critical for the chalcopyrites class, especially in addressing the topological properties\citep{Surasree2022} and exploring its dynamical stability and thermodynamic characteristics is of utmost importance. We have calculated the phonon density of states and confirmed the absence of any negative frequency phonon DOS. The negative Helmholtz free energy as a function of temperature and positive entropy, depicted in the inset of Fig. S1(c), further supports the thermodynamic stability of  HgSnN$_2$. Similarly, Fig. S1(f) illustrates the phonon density of states without any negative frequency region and the temperature-dependent positive evolution of Helmholtz free energy for HgSnP$_2$. These calculations assure the stability of our structure from a thermodynamic standpoint. 
\par
\textcolor{black}{Additionally, we have depicted the phonon dispersion of HgSnN$_2$ and HgSnP$_2$ in Fig. S2(a) and Fig. S2(b), respectively, with two distinct modes—acoustic and optical—i.e., respective the lowest and the highest frequencies observed at the $\Gamma$ point in the phonon dispersion. To enhance comprehension of both optical and acoustic modes, we have provided a visual representation of displacement vectors within the primitive cell arrangement. In Fig. S2(a), the acoustic mode registers a frequency of 0.016 THz, while the optical mode is observed at 18.97 THz. Interestingly, no phonon bands are within the frequency range of 6.48 THz to 8.80 THz. Examining the acoustic mode of HgSnN$_2$, we note that mainly cations (Hg and Sn atoms) exhibit prominent displacement vectors along the negative X and positive Y directions, with no displacement in the Z-direction. The specific components of displacement vectors associated with each atom are detailed in Table \ref{Acoustic}, and the corresponding displacement vector is illustrated in Fig. S2(c). Conversely, in the optical mode, cations (Hg and Sn atoms) display no displacement vector. Within this mode, anions (N) predominantly contribute to the vibrations. The detailed components of displacement vectors are listed in Table \ref{Optical}, and the corresponding displacement vector is depicted in Fig. S2(d).}

\textcolor{black}{In the phonon dispersion of HgSnP$_2$, illustrated in Fig. S2(b), the acoustic mode manifests at a frequency of 0.010 THz, while the optical mode is discerned at 9.71 THz. Notably, there are no phonon bands within the frequency range of 4.52 THz to 7.20 THz. Upon closer examination of the acoustic mode in HgSnP$_2$, it is observed that cations (Hg and Sn atoms) prominently exhibit displacement vectors along the Z-direction with tiny displacement of anions (P). The specific components of these displacement vectors are comprehensively outlined in  Table \ref{Acoustic}, with the corresponding visual representation of the displacement vector provided in Fig. S2(e). Conversely, in the optical mode, cations (mainly Sn atoms) showcase relatively less displacement vector compared to anions (P). Within this mode, anions (P) play a predominant role in contributing to the vibrations dynamics. The intricate details of the displacement vectors are cataloged in Table \ref{Optical}, and the corresponding displacement vector is portrayed in Fig. S2(f).}
\textcolor{black}{The notable resemblance between the two sibling compounds lies in the dominance of cations (Hg and Sn) contributions in the acoustic mode. Also, in the optical mode, the participation is primarily from anions, with X (where X = N/P) playing a major role. A significant disparity is observed in the displacement vectors in the acoustic mode for HgSnN$_2$ Hg, and Sn shows displacement along X and Y directions, but for HgSnP$_2$, it is in the Z-direction. Another crucial observation is Sn has taken part in optical mode only for HgSnP$_2$.} 
\textcolor{black}{We have also determined the elastic stiffness constants by assessing the energy variation by applying minor strains to the equilibrium lattice configuration to ensure the elastic stability of HgSnX$_2$ (X=N/P). The count of elastic stiffness constants increases as the symmetry of the crystal structure is reduced. In the case of our system, which belongs to the tetragonal group, there are six independent elastic stiffness constants namely   C$_{11}$, C$_{12}$, C$_{13}$, C$_{33}$, C$_{44}$, C$_{66}$\citep{mouhat2014necessary,baghdad2022study}. The value of the elastic constant is tabulated in Table \ref{Elastic}. A material is considered elastically stable when the elastic constants meet the criteria (1) to (4) for a tetragonal structure with the space group 122. We have found that both compounds satisfy the four necessary and sufficient conditions, which ensures the mechanical stability of these two compounds.}
\begin{equation}
\textcolor{black}{C_{11} > |C_{12}|}
\label{E4}
\end{equation}
\begin{equation}
\textcolor{black}{
(C_{11}+C_{12})C_{33} > 2(C_{13})^2}
\label{E4}
\end{equation}
\begin{equation}
\textcolor{black}{C_{44} > 0.0}
\label{E4}
\end{equation}
\begin{equation}
\textcolor{black}{C_{66} > 0.0}
\label{E4}
\end{equation}


\begin{table*}
\centering
\begin{tabular}{|c|c|c|c|c|c|c|c|}
\hline
Material & C$_{11}$ (GPa) & C$_{12}$ (GPa) & C$_{13}$ (GPa) &  C$_{33}$ (GPa) &  C$_{44}$ (GPa) &  C$_{66}$ (GPa)\\
 
\hline
\textcolor{black}{HgSnN$_2$}&\textcolor{black}{206.294}&\textcolor{black}{74.055}&\textcolor{black}{64.916}&\textcolor{black}{223.207}&\textcolor{black}{101.362}&\textcolor{black}{48.705}\\
\hline
\textcolor{black}{HgSnP$_2$}   &\textcolor{black}{338.194} & \textcolor{black}{-26.859} & \textcolor{black}{-54.115} & \textcolor{black}{124.782} & \textcolor{black}{57.694} & \textcolor{black}{30.692} \\
   \hline              
\end{tabular}
\caption{\textcolor{black}{The six independent  elastic stiffness constants for HgSnN$_2$ and HgSnP$_2$ shows the elastic stability of the materials.}}
\label{Elastic}
\end{table*}

\section{Calculation methodology}
We conducted electronic structure calculations based on Density Functional Theory (DFT) using the Vienna $\textit{ab-initio}$ simulation package (VASP) with a plane-wave basis set and pseudo-potential framework\citep{vasp1, vasp2}. The calculations employed the Perdew–Burke–Ernzerhof (PBE) prescription, which is a generalized gradient exchange-correlation approximation (GGA) functional\citep{PBE}. The influence of SOC was considered by applying a relativistic correction to the original Hamiltonian. A cut-off energy of 600 eV was used for the plane-wave basis. We relaxed the internal atomic positions to optimize the structure until the Hellman–Feynman force was less than 0.001 eV/$\AA$. The Brillouin zone was sampled using an 8 $\times$ 8 $\times$ 6  k-point mesh, and the electronic convergence criteria were set at 10$^{-9}$ eV for self-consistent calculations. We constructed a tight-binding model using the Wannier function basis set to study the topological properties. This energy-selective method mainly addresses the low-energy, few-band model Hamiltonian defined in the effective Wannier function basis. Using a re-normalization method, we integrated out the irrelevant degrees of freedom to keep the orbitals that are solely responsible for topological effects. In the down-folding calculations, we considered only the \textcolor{black}{Hg-${5d}$, Sn-${5p}$, X-${np}$, X-${ns}$ (n=2,3 for X=N/P)} orbitals as active degrees of freedom for HgSnN$_2$ and HgSnP$_2$ respectively, while other degrees of freedom were down-folded as they made less significant contributions in determining the topological properties. To analyze the topological aspects, we calculated the topological index (Z$_2$ number), examined the surface states, studied the distribution of normalized Berry curvature in the 2D momentum plane, and investigated the Fermi arc at the particular chemical potential of Weyl crossing. These calculations were performed using Wannier90\citep{ wannier1, wannier2, wannier3} and WannierTool\citep{wanniertool}, starting from the full converged self-consistent DFT outputs.

\section{Electronic structure}
\begin{figure*}
\includegraphics[width=17cm]{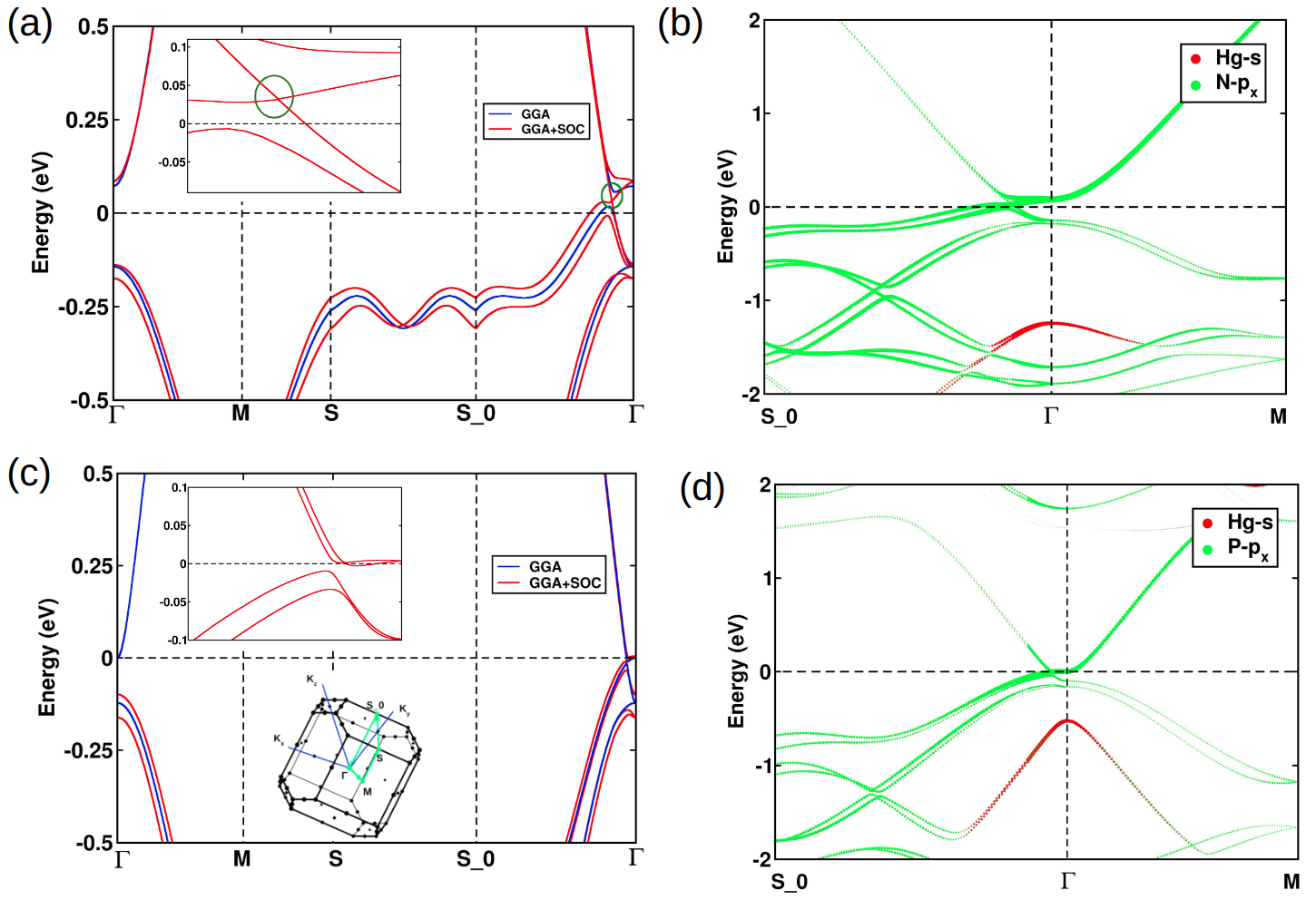}
\caption{(Color online) (a) and (c) depict the calculated bulk electronic band structures for HgSnN$_2$ and HgSnP$_2$, respectively, employing the GGA (blue lines) and GGA+SOC (red lines). The inset images in (a) and (c) provide a zoom version of the gap-less and gaped states in the GGA+SOC band structures for HgSnN$_2$ and HgSnP$_2$, respectively. (b) and (d) illustrate the calculated orbital projected band structures with GGA+SOC for the Hg and X (X=N, P) atoms, showing the band inversion.\textcolor{black}{ The Brillouin zone (BZ) is shown in the lower inset of (c).} The energy scale is referenced to the Fermi energy(E$_f$), which is set at zero.}

\label{Fig-2-DFT-bands}
\end{figure*}
The GGA density of states (DOS) for HgSnX$_2$ (X=N, P) in the nonmagnetic state has been successfully determined using first-principles density functional theory, shown in Fig. \ref{Fig-1-Struc-DOS}(b)-(c). By analyzing the DOS of HgSnN$_2$ in Fig.\ref{Fig-1-Struc-DOS}(b), it becomes evident that it exhibits a semi-metallic nature due to the absence of a noticeable gap in DOS at the Fermi energy, primarily influenced by the significant $p$-orbital contribution of N. However, upon replacing N with P, the contribution from the $p$-orbital of P shifts slightly away from the Fermi energy, as depicted in Fig. \ref{Fig-1-Struc-DOS}(c), opening a small gap at Fermi energy in DOS. In both cases, the DOS shows that the main contributions near the Fermi energy originate from \textcolor{black}{{Hg-${5d}$}}, \textcolor{black}{Sn-${5p}$, and X-${np}$ (n=2,3 for X=N/P)} orbitals, with minor contributions from  \textcolor{black}{Sn-$5s$, and X-$ns$ (n=2,3 for X=N/P)} orbitals. However, these minor parts are not shown in the DOS plots. Notably, X-$np$ (n=2,3 for X=N/P) orbitals dominate over \textcolor{black}{Hg-${5d}$} orbitals near the Fermi energy, and the presence or absence of states in DOS near the Fermi energy is mostly determined by the $p$-orbital contribution from the X atom, which is further crucial for understanding the electronic band structure.

To gain a comprehensive understanding of the topological aspects, we have analyzed the electronic band structure of HgSnX$_2$ (X=N, P). We have plotted the band structures of HgSnX$_2$ (X=N, P) using both the GGA and GGA+SOC methods, focusing on the few high symmetry points, as shown in Fig. \ref{Fig-2-DFT-bands}. A typical connection of both bands is the removal of degeneracy in the GGA bands in the presence of SOC, which leads to the involvement of these bands in generating nontrivial topological signatures and forming a gapless state (band crossing) or band gap. Specifically, in Fig. \ref{Fig-2-DFT-bands}(a), we have noticed that the incorporation of SOC has a fascinating effect on the degeneracy of the GGA band structure. SOC has broken the two-fold bands in both the valence and conduction bands near Fermi energy. Subsequently, a noteworthy occurrence takes place when a particular section of those valence and conduction bands come close to each other and again form a point degeneracy in the band structure, which leads to Weyl crossing protected by two-fold degeneracy at a non-high symmetry point in momentum space. These coordinates are found to be (-0.05534, 0.07620, -0.04912) along the path from S\textunderscore0 to $\Gamma$. It is important to highlight that the significant role played by SOC is the creation of point degenerate Weyl crossing for HgSnN$_2$, while the gapped band structure for HgSnP$_2$. 

Additionally, by examining the orbital-projected bands, we have found that the major contribution in the Hg atom (like $\Gamma_6$) comes from the Hg-$6s$ orbital, which is lower in energy than the major contribution from the N-$2p$ orbital (like $\Gamma_8$). This observation leads to a negative band gap at the $\Gamma$ symmetry point, supporting a nontrivial topological phase. We have also observed a contribution from the N-$2s$ orbital in addition to the Hg-$6s$ orbital, as depicted in Fig. S3(a). The $p$ orbitals of N (2p$_y$ and 2p$_z$) are also marked in Fig. S3(b) and Fig. S3(c), respectively. Therefore, conducting a detailed investigation of the band-crossing phenomenon is crucial in light of the relevant topological signatures. In the case of HgSnP$_2$, an interesting observation was made when incorporating the SOC strength. It led to the elimination of the two-fold band degeneracy in both the valence and conduction bands; consequently, the specific regions of these bands come close to each other, resembling the behavior observed in HgSnN$_2$. However, unlike HgSnN$_2$, the bands in HgSnP$_2$ do not exhibit a Weyl crossing. Instead, a notable distinction was noticed where a global band gap of approximately 14 meV emerges between the valence and conduction bands at the  $\Gamma$ point, as illustrated in Fig. \ref{Fig-2-DFT-bands}(c). There is still band inversion at the high symmetry $\Gamma$ point, where the major contribution comes from the s orbitals of Hg and P ($\Gamma_6$), which are lower in energy than the major p orbital contribution ($\Gamma_8$), as shown in Fig. \ref{Fig-2-DFT-bands}(d), Fig. S3(d), Fig. S3(e), and Fig. S3(f). The point to be noted is that it is essential to acknowledge that the presence of SOC primarily alters the characteristics of the GGA bands. The introduction of SOC leads to the breaking of degeneracy in the GGA bands for both compounds, resulting in a state with either no energy gap or a noticeable gap in the GGA+SOC bands. This intriguing observation prompts us to conduct a more in-depth analysis of the topological characteristics exhibited by both compounds and, simultaneously, to see the impact of SOC on their respective topological phases. Interestingly, although the strength of SOC is identical in both compounds, but they still display distinct topological phases. Therefore, in the next section, our investigation aims to uncover the underlying reasons behind this disparity, and the anticipation regarding the role of SOC also demands further exploration.


\section{Topological investigation}

In this section, we have extensively examined the topological properties of both compounds by implementing an \textit{ab-initio}-derived model Hamiltonian approach. Our main objective was to discern and understand the distinctive topological traits displayed by these materials. 


\subsection{Topological signatures of HgSnN$_2$ }

\begin{figure*}
\includegraphics[width=17cm]{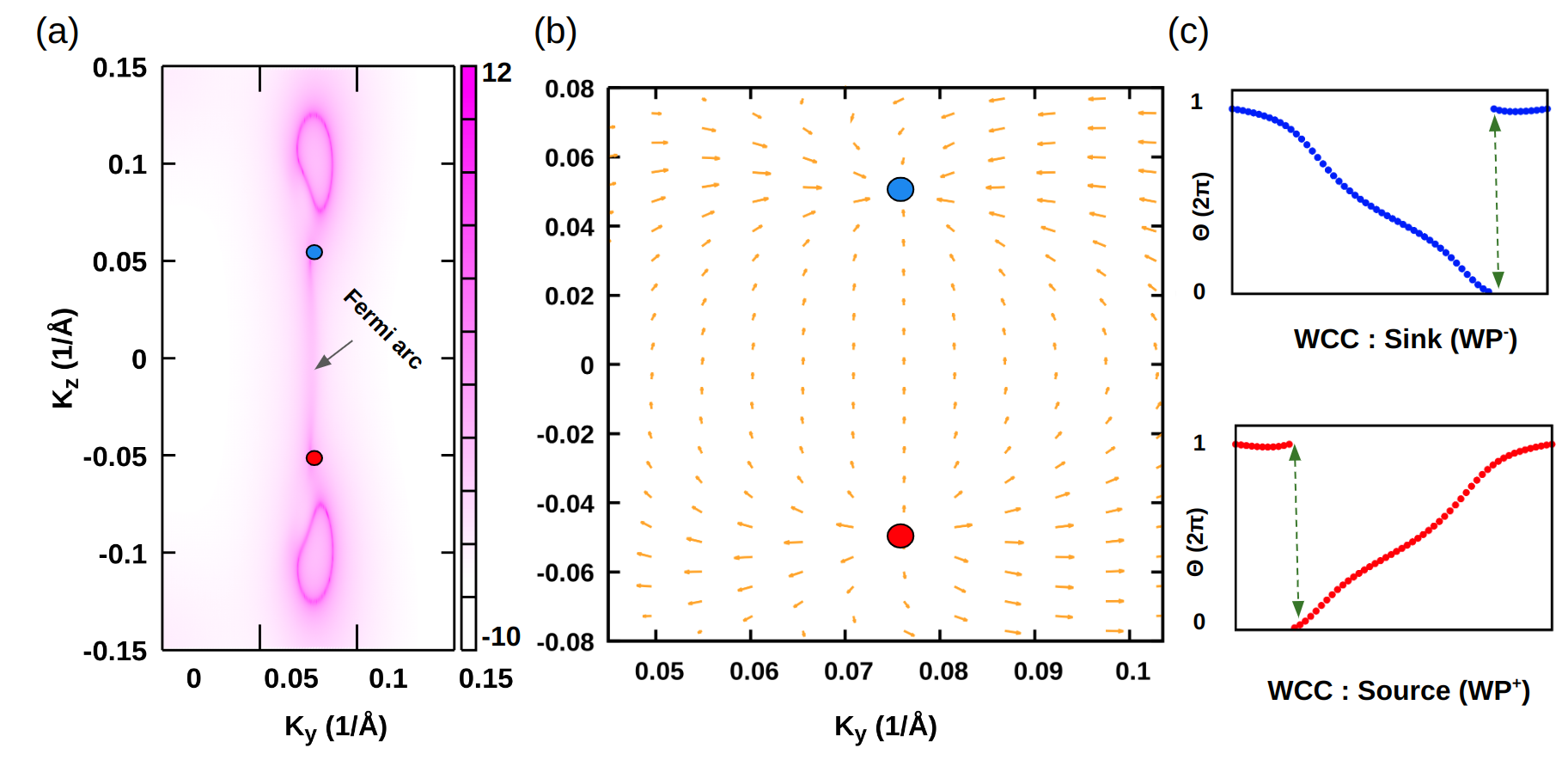}
\caption{(Color Online) (a) The calculated Fermi arcs, with a chemical potential positioned approximately 7 meV above the Fermi energy (E$_f$), establish connections between the opposite chiral Weyl points in the momentum space. These Weyl points are indicated by red (source) and blue (sink) filled circles in the $K_y-K_z$ momentum plane. In the same momentum plane, (b) illustrates the corresponding distribution of normalized Berry curvatures in the outward (source) and inward (sink) direction of fluxes between the two opposite chiral Weyl points. Displayed in (c) is the flow of the Wannier Charge Centers (WCC) that move in opposite directions for source and sink points. }
\label{Fig-3-fermiarc}
\end{figure*}

In HgSnN$_2$, we have successfully identified a band crossing phenomenon in the band structure along the S\textunderscore0 to $\Gamma$ path, as illustrated in Fig. \ref{Fig-2-DFT-bands}(a), which exhibits all the characteristic features of a time-reversal symmetry protected Weyl semimetallic phase. In a Weyl semimetal, there exists a Fermi arc connecting two Weyl crossings of opposite chirality in momentum space. These Weyl crossings occur in pairs and can be viewed as magnetic monopoles in momentum space. For HgSnN$_2$, we discovered a similar outcome that two Weyl crossings, WP+ with a momentum space coordinate of (-0.0553, 0.0762, -0.0491) and WP- with a coordinate of (-0.0550, 0.0759, 0.0493) which are connected by a Fermi arc in the $K_y-K_z$ momentum plane, as depicted in Fig. \ref{Fig-3-fermiarc}(a), specifically around an energy contour close to the chemical potential, approximately 7 meV above the Fermi energy level. The opposite chirality of the Weyl crossings is clearly evident from their normalized Berry curvature distribution, as displayed in Fig. \ref{Fig-3-fermiarc}(b) in the same momentum plane. The Berry curvature near the blue-filled circle is inward, acting as a sink in momentum space, while the filled red circle represents a source with an outward field line direction. To validate our analysis further, we examined the evolution of the Wannier Charge Center (WCC) throughout the Brillouin Zone (BZ). Our analysis revealed that the average position of the WCC, calculated using the Wilson-loop method, shows a discontinuous jump marked by a double-headed arrow. Notably, the average position of the WCC for the sink (upper panel) is in the opposite direction compared to the source (lower panel). We have enlisted the coordinates of the Weyl crossing with its chirality and energy positions with respect to the Fermi level in Table V.

\begin{table}[htbp]
\renewcommand*{\arraystretch}{1.2}
    \centering
    	\begin{tabular}{|c|c c c|c|c|} 
\hline             
            & \multicolumn{3}{|c|}{Coordinates} & Chemical & Chirality \\ 
   Compound     & \multicolumn{3}{|c|}{in K-space (1/$\AA$)} & Potential &  \\
            & \multicolumn{3}{|c|}{} & (meV) & \\
            \hline 
 HgSnN$_2$   & -0.0550 &  0.0759 &  0.0493 &  6.8  & -1 (WP-) \\
                            &  -0.0553 &  0.0762 & -0.0491 &  7.9 & +1 (WP+)\\
\hline         
\end{tabular}
\caption{In the momentum space, the Weyl node coordinates (in unit of 1/$\AA$) of HgSnN$_2$ with 
their chemical potentials relative to the Fermi energy, and chiralities are listed.}
\label{Weyl-table1}
\end{table}

We have also calculated the spectral function using the slab model approach as implemented in 
WannierTool\citep{wanniertool} and found that two Weyl points are connected by a Fermi arc-
related band in the surface state, as shown in supplementary  Fig. S5(a). Collectively, all these 
findings in HgSnN$_2$ firmly establish it as a nonmagnetic Weyl semi-metal, possessing all the key 
topological characteristics. Due to the proximity of Weyl crossing to the Fermi energy, this 
compound can be regarded as an excellent candidate for an Hg-based ideal Weyl semi-metal 
\citep{ruan2016ideal}, making it highly suitable for experimental investigations of the Fermi 
arc. Furthermore, the Weyl crossing observed in this compound is sufficiently distinct and 
isolated from the surrounding bulk bands, facilitating experimental observations and analysis.

\subsection{Topological signatures of HgSnP$_2$ }
\begin{figure}
\center

\includegraphics[scale=0.6]{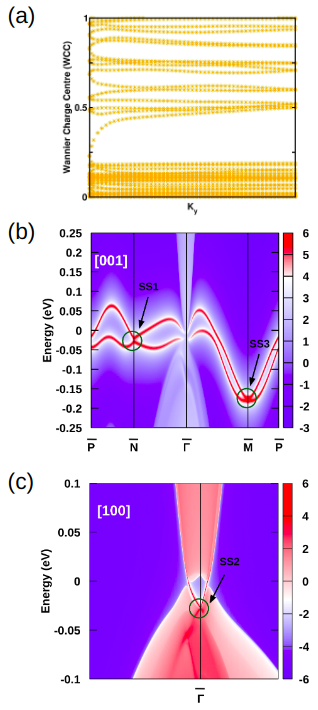}
\caption{(Color online) (a) Depicts the flow of average Wannier Charge Centers (WCC) position for the highest occupied valence band of HgSnP$_2$, obtained through the Wilson-loop method. Notably, the flow showcases a continuous trajectory of the WCC. (b) The surface spectral function across the surface Brillouin zone (BZ) is represented in this figure for HgSnP$_2$. A green circle highlights the presence of the odd numbers of the Dirac crossing. \textcolor{black}{(c) shows the Dirac cone at  $\Gamma$ point shown along [100] for clear visualization.} The energy scale is referenced to the Fermi energy(E$_f$), which is set at zero.}

\label{Fig-4-HgSnP2}
\end{figure}
The presence of a narrow band gap of 14 meV and a band inversion between the $\Gamma_6$ and $\Gamma_8$ bands, as depicted in Fig. \ref{Fig-2-DFT-bands}(c) and Fig. \ref{Fig-2-DFT-bands}(d), prompted us to investigate the topological properties of HgSnP$_2$. While we did not observe any band crossings similar to its sister compound HgSnN$_2$, but the existence of band inversion with a small band gap between the maxima of valence and minima of the conduction band is a significant indication that could be categorized as a topological insulator. Following standard practices for studying topological behaviors, we examined the evolution of the WCC along the K$_y$ direction of the Brillouin Zone (BZ) in K$_x$=0.0 plane, as shown in Fig. \ref{Fig-4-HgSnP2}(a). We found that the WCC exhibits continuity, indicative of its nontrivial topological nature. Additionally, we plotted the same quantities, i.e., WCC, along different directions in the BZ plane, as shown in the supplementary Fig. S5(b) other momentum planes in the K$_x$=0.5, K$_y$=0.5, and K$_z$=0.5 which are discontinuous. Based on these findings, we conclude that HgSnP$_2$ possesses a nonzero topological index with a $Z_2$ number [1,000] which falls under the category of strong topological insulator.

\par

\textcolor{black}{To further support our conclusions, we calculated the spectral function or surface state in the [001] and [100] direction, scanning the entire BZ at time-reversal invariant momenta (TRIM) points. The analysis revealed a gap-less surface state with an odd number of Dirac crossings, as illustrated in Fig. \ref{Fig-4-HgSnP2}(b)-(c). We analyzed the spectral distribution at the surface of the semi-infinite slab of HgSnP$_2$. The presence of a Dirac-type gap-less crossing in the spectral function serves as a crucial indicator for identifying the topological state. This observation aligns with our earlier findings related to bulk bands, affirming the consistency of our results. We utilized Green's function techniques, employing the WannierTool\citep{wanniertool} to extract the tight-binding model through maximally projected Wannier functions (WFs). In our visual representation, the red bands correspond to surface bands against a blue background of bulk bands. For example, no crossing is observed at the M high symmetry point of the bulk band, approximately -0.15 eV. However, in the surface bands, we identified a Dirac-like band crossing. These crossings are denoted as SS1, SS2, and SS3, with SS1 located at the N high symmetry point, SS2 at the $\Gamma$ high symmetry point, and SS3 at the M high symmetry point along the [001] and [100] direction. Since SS1 is not very clear along the [001] direction, we have identified it clearly in the [100] directions. We have put an image along [100] to show the Dirac cone at  $\Gamma$ point for clear visualization in Fig. \ref{Fig-4-HgSnP2}(c). Consequently, our analysis indicates an odd number of Dirac cones in the surface state, providing valuable insights into the topological characteristics of the system.} 
\par
The $Z_2$ number [1,000] and the existence of an odd number of Dirac cones unequivocally establish HgSnP$_2$ as a strong topological insulator, characterized by a bulk band gap and a metallic surface state protected by time-reversal symmetry. We have also observed spin momentum locking, as shown in supplementary Fig. S5(c). While the chalcopyrite class offers numerous examples of topological insulators, our current understanding suggests a lack of comprehensive studies focusing on the topological aspects of Hg-based chalcopyrites in the existing literature. In a nutshell, the relevant signature indicates that HgSnP$_2$ is a strong topological insulator.

\section{Effect of SOC modulation}

\begin{figure*}
\includegraphics[width=17cm]{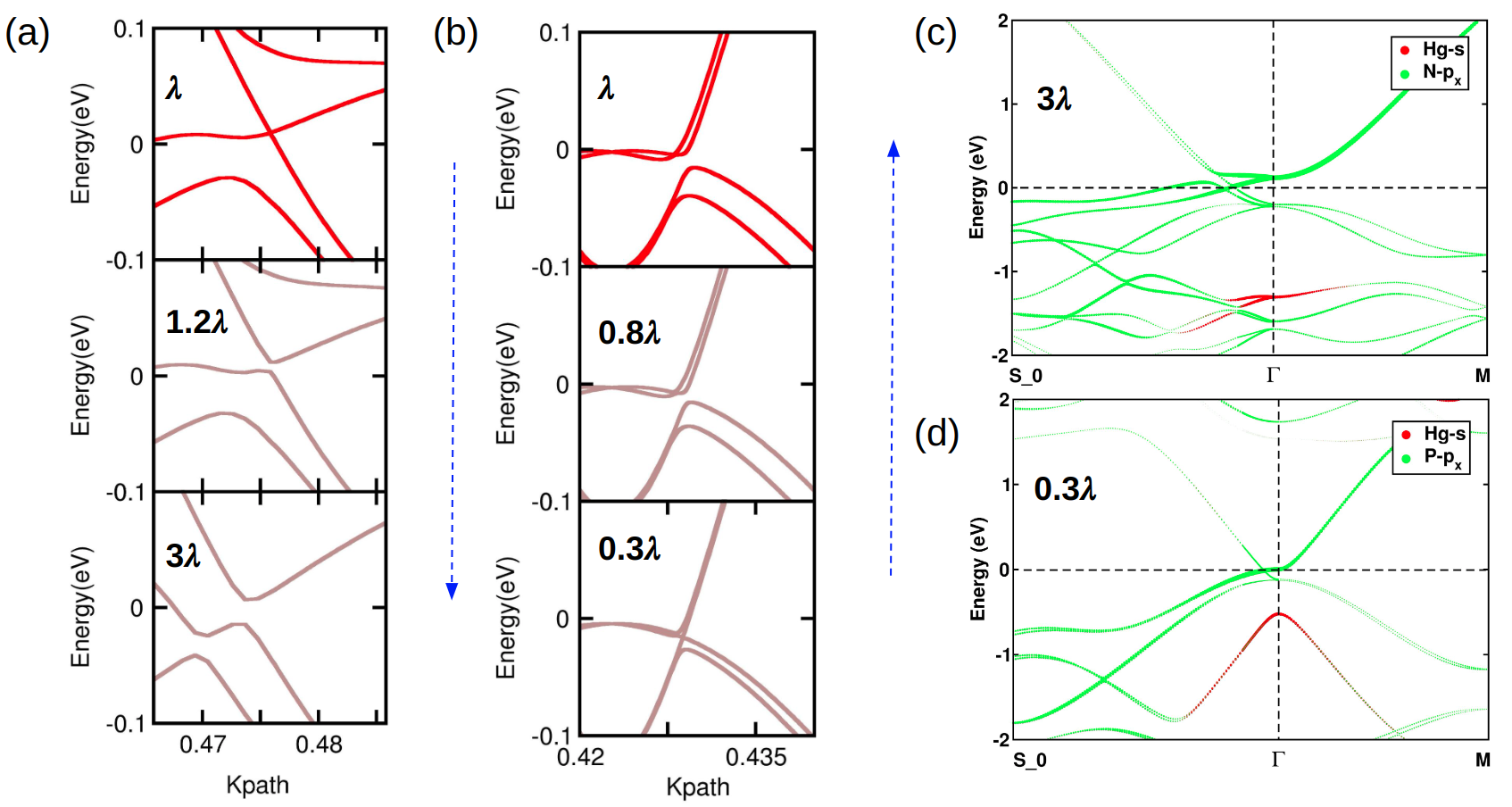}
\caption{(Color online) Calculated GGA+SOC band structure for (a) HgSnN$_2$ and (b) HgSnP$_2$ with different SOC strength ($\lambda$). (c) and (d) represent the projected band structures which show the band inversion at 3$\lambda$ ($\xi$=3) and 0.3 $\lambda$ ($\xi$=0.3) for HgSnN$_2$ and HgSnP$_2$ respectively. Here $\lambda$ is the original SOC strength, and $\xi$ is the scaling factor of SOC. The energy scale is referenced to the Fermi energy(E$_f$), which is set at zero.}
\label{Fig-6-SOC}
\end{figure*}

The primary focus revolves around comprehending the fundamental factors that give rise to the distinct (gap and gap less) topological phases observed in two closely related compounds, namely, HgSnN$_2$ and HgSnP$_2$, which have the same chemical composition except for the anion (N/P). These compounds share the same space group with broken inversion symmetry and preserving time-reversal symmetry. In this context, it is essential to emphasize that Hg has strong atomic SOC strength, while atomic N or P does not contribute to SOC strength significantly. Interestingly, despite both compounds having the same atomic SOC strength, but they still display different topological phases, which prompts us to delve deeper into the underlying reasons behind this non-identical behavior.

To reinforce our analysis further, we have taken into account a $\xi$ term in the relativistic Hamiltonian, H$_{\text{SOC}}$ = $\xi$$\lambda$ $\Vec{L}$.$\Vec{S}$, where $\Vec{L}$ and $\Vec{S}$ represent orbital and spin angular momentum, respectively as followed by the prescription\citep{kanungo2016weak,kanungo2022SOC} and $\lambda$ is the original SOC strength while $\xi$ serve as a modulating factor of original SOC strength. \textcolor{black}{Generally, SOC is an inherent atomic property of the element that remains unaltered with the same nominal valance state. However, we can artificially tune the SOC by scaling the relativistic term in the DFT calculations. Indeed, the artificial increase or decrease of SOC is not factually correct and does not represent the physical picture. However, such model calculations will be very effective in understanding the impact of the SOC in comparison to the other energy scales involved in the materials.}
In the case of original SOC strength $\xi$=1, but we can arbitrarily vary the value of $\xi$ to probe its effect. We have explored the consequences by sequentially scaling up and down the value of $\xi$ in Hg-based materials. The strength of the SOC is dependent on the fourth power of the atomic number, meaning that the effect of SOC becomes more dominant as we move down the periodic table and very strong for Hg. For HgSnN$_2$, we have utilized values of 1.2$\lambda$ (i.e $\xi$=1.2) and 3$\lambda$ (i.e $\xi$=3), while for HgSnP$_2$, the effect has been reduced to 0.8$\lambda$ (i.e $\xi$=0.8) and 0.3$\lambda$ (i.e {$\xi$}=0.3). Remarkably, we have observed that we can artificially calibrate multiple topological phases by scaling up and down {$\xi$} parameter in both materials.

Now, we investigated the effect of varying SOC strength on the Weyl crossing in HgSnN$_2$. In Fig. \ref{Fig-6-SOC}(a), we increased the strength of SOC ($\xi$), say,  up to three times the original SOC strength (3$\lambda$) and observed that at 1.2$\lambda$, the gapless feature disappears, and a small gap begins to form, which becomes more prominent at 3$\lambda$. These two artificial SOC-driven band structures at 1.2$\lambda$ and 3$\lambda$ in HgSnN$_2$ indicate that relativistic SOC indeed plays a crucial role in maintaining the gapless nature of the band structure. We performed a similar analysis on HgSnP$_2$ for cross-validation, as shown in Fig. \ref{Fig-6-SOC}(b). In this case, we reduced the strength of SOC, say, to 0.8$\lambda$ and 0.3$\lambda$. We found that the band gap at the same momentum coordinate vanishes. At 0.8$\lambda$, the band gap between the valence and conduction bands is reduced; at 0.3$\lambda$, it becomes gap-less. However, it is essential to note that the band structure at 0.3$\lambda$ in HgSnP$_2$ is already artificially influenced in terms of SOC, so further verification of its topological characteristics is not meaningful. However, from first-principles DFT calculations, we can verify the nontrivial properties by observing the band ordering in terms of orbital contributions, whether they are normal or inverted. We observed that in HgSnN$_2$ at SOC strength 3$\lambda$, the major Hg-$6s$ orbital contribution ($\Gamma_6$) is lower than the major N-$p$ orbital contribution ($\Gamma_8$) as shown in  Fig. \ref{Fig-6-SOC}(c). Furthermore, the involvement of the N-$2s$ orbital has been identified alongside the Hg-$6s$ orbital, illustrated in Fig. S4(a). The N-$2p$ orbitals (2p$_y$ and 2p$_z$) have also been designated in Fig. S4(b) and Fig. S4(c) correspondingly.
Hence, a negative band gap is present at the high-symmetry point ($\Gamma$), similar to the original band structure as shown in Fig. \ref{Fig-2-DFT-bands}(b).
Similarly, in HgSnP$_2$ at 0.3$\lambda$, the major Hg-$6s$ orbital contribution from the Hg atom 
($\Gamma_6$) is lower than the major P-$3p$ orbital contribution ($\Gamma_8$) as shown in  
Fig. \ref{Fig-6-SOC}(d). We have observed an additional involvement of the P-$3s$ orbital alongside the Hg-$6s$ orbital, which is illustrated in Fig. S4(d). The $3p$ orbitals of P, namely 3p$_y$ and 3p$_z$, have also been indicated in Fig. S4(e) and Fig. S4(f), respectively.
Notably, both band structures at artificially modified SOC strengths also exhibit band inversion. The study of these sister compounds enables us to examine the role of SOC 
in driving the topological phase in quantum materials from a modeling perspective. Another 
critical perspective is that we have noticed a striking resemblance between the gapped band 
structure of HgSnN$_2$ at 0.3$\lambda$ and the parent band structure of HgSnP$_2$ at its original 
SOC strength. Similarly, a comparable equivalence can be established for HgSnP$_2$, where the 
gapless band structure at 3$\lambda$ closely resembles the parent state of HgSnN$_2$ at its 
original SOC strength. In summary, by modulating the SOC strength, we observe the crossover from gapped to gapless and vice-versa topological phases can emerge in these two sister compounds, which can also be achievable by tuning the chemical composition via N or P. 

\begin{figure}
\center
    \includegraphics[width=\linewidth]{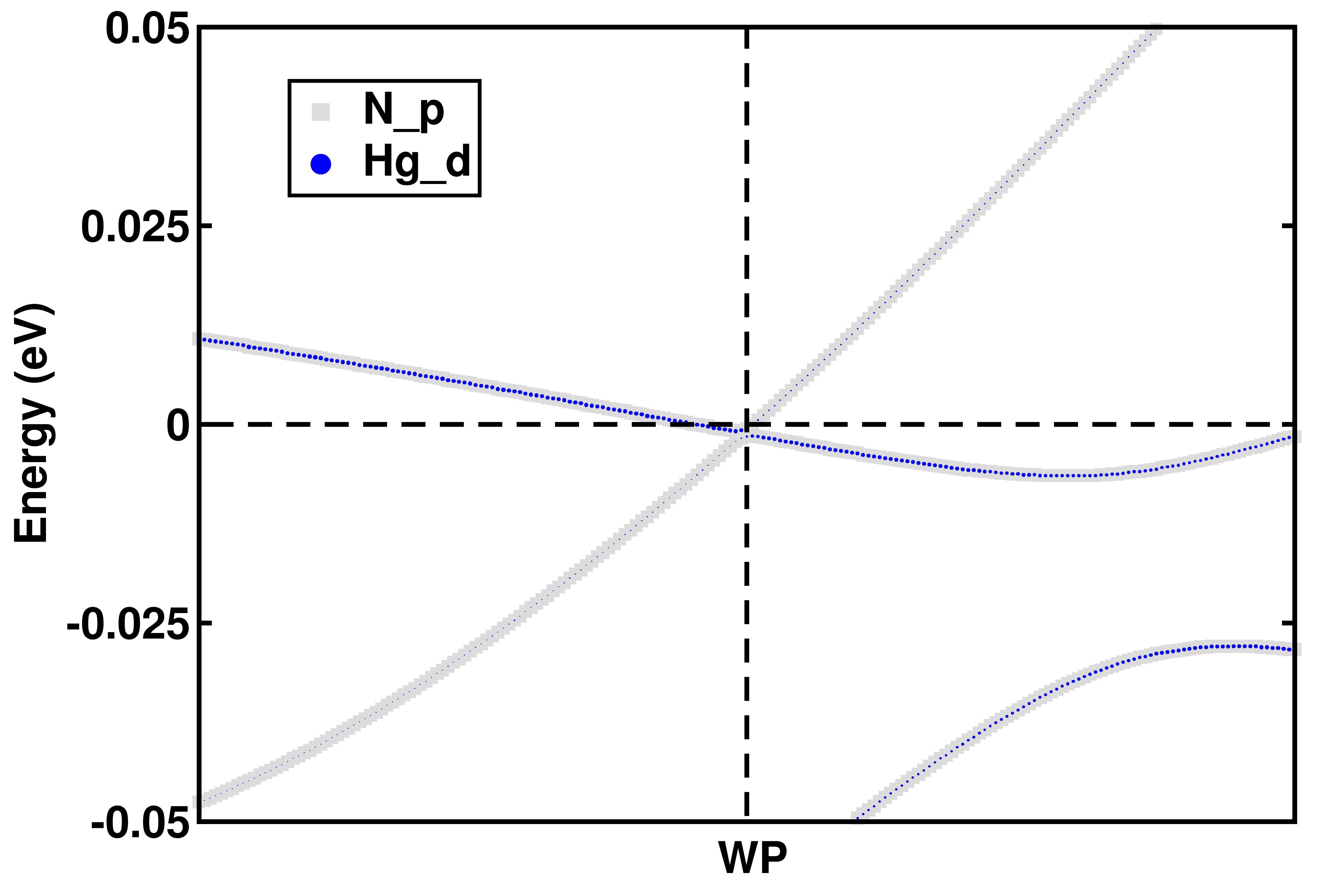}
    \caption{(Color online)The GGA+SOC orbital projected hybridized band structure of HgSnN$_2$  with major  N-$p$ (grey) and minor \textcolor{black}{Hg-${5d}$}(blue) orbitals. The energy scale is referenced to the Fermi energy(E$_f$), which is set at zero (E$_f$=0.0 eV) }
\label{Fig-5-Hg-d}
\end{figure}

Ideally, both the compound HgSnN$_2$ and HgSnP$_2$ should be the same in terms of SOC. To correlate such a strong effect of SOC in these compounds, we found a notable feature that the band crossing or Weyl crossing observed in HgSnN$_2$ mainly arises from the hybridization of the major N-$p$ orbital and minor \textcolor{black}{Hg-${5d}$} orbital states, as shown in Fig. \ref{Fig-5-Hg-d}. This hybridization plays a crucial role in explaining the topological characteristics of the respective compounds. The SOC strength is particularly effective in the hybridized band due to the presence of \textcolor{black}{Hg-${5d}$} orbitals. The minor contribution from \textcolor{black}{Hg-${5d}$} orbitals may modify the band dispersion without substantial changes in the overall band structure. The impact of SOC on the topological phase is widely recognized, but this study has elegantly illuminated the crucial role of hybridization in modifications to the topological phase.


 \section{Summary and Conclusion}

In this paper, we have undertaken a comprehensive \textit{ab-initio}  analysis of two topological materials, namely HgSnN$_2$ and HgSnP$_2$, highlighting their intriguing topological properties driven by SOC. We confirmed the structural stability of the compounds by the absence of phonon-negative frequency modes. In our investigation, we found that HgSnN$_2$ exhibits characteristics of a nonmagnetic Weyl semi-metal, and the Weyl crossing is protected by a point band degeneracy. In the case of HgSnP$_2$, we found a similar band inversion even after the N is replaced by P. This change in chemical composition effects the topological properties, mainly opening a small band gap. In particular, HgSnP$_2$ is identified as a strong topological insulator. The point to be noted is that although both the compounds offer the same atomic SOC strength, but the hybridization of bands between the elements present in our compounds plays a critical role in deciding whether the topological phase is a topological insulator or semi-metal.  Furthermore, we have observed that by adjusting the strength of SOC in the model calculation, for the case of  HgSnN$_2$, it is possible to generate equivalent topological signatures to those of the parent HgSnP$_2$, and vice versa. Such tuning of SOC can be accomplished through doping, substitution, and the design of heterostructures in the quantum materials\citep{dziawa2012topological,mahfouzi2012spin,naimer2023twist}.



Exploring topological properties in Hg-based chalcopyrite has been relatively less reported compared to other chalcopyrites. We have observed that chalcopyrites based on N and P have been successfully synthesized, including the formation of single crystals\cite {zawilski2010growth,vohl1979synthesis}. In contrast, both structures lack inversion symmetry but preserve time-reversal symmetry, making our study simple to capture the dependence of SOC effectively. This type of work can be further extended for any suitable choice of such sibling magnetic compounds to capture the interplay of topology and magnetism.  Furthermore, the remarkable proximity of the Weyl crossings to the Fermi energy in the case of HgSnN$_2$ suggests the potential for experimental detection.  These findings pave the way for exciting advancements in these fields.\\

\section{Acknowledgments}
SS expresses gratitude to IIT Goa for providing the research fellowship under the Government of India.


\begin{thebibliography}{99}

\bibitem{krasovskii2015spin} E E Krasovskii, J. Condens. Matter Phys. $\bf{27}$, 493001 (2015).
\bibitem{fan2014quantifying} X. Fan, H. Celik, J. Wu, C. Ni, K. J. Lee, V. O. Lorenz and J. Q. Xiao, Nat. Commun. $\bf{5}$, 3042 (2014).
\bibitem{soumyanarayanan2016emergent} A. Soumyanarayanan, N. Reyren, A. Fert and C. Panagopoulos, Nature $\bf{539}$, 509 (2016).
\bibitem{moon2013non} E. G. Moon, C. Xu, Y. B. Kim and L. Balents, Phys. Rev. Lett. $\bf{111}$, 206401 (2013).
\bibitem{huang2017interplay} B. Huang, Y. B. Kim, and Y. M. Lu, Phys. Rev. B $\bf{95}$, 054404 (2017).
\bibitem{catuneanu2018path} B. Huang, Y. B. Kim, and Y. M. Lu, npj Quantum Mater. $\bf{3}$, 23 (2018).
\bibitem{kitagawa2018spin} K. Kitagawa, T. Takayama, Y. Matsumoto, A. Kato, R. Takano, Y. Kishimoto, S. Bette, R. Dinnebier, G. Jackeli and H. Takagi,  $\bf{554}$, 341 (2018).
\bibitem{kim2015kitaev} H. S. Kim, V. Shankar V., A. Catuneanu, and H. Y. Kee, Phys. Rev. B $\bf{91}$, 241110 (2015).
\bibitem{shick2019spin} A. B. Shick and W. E. Pickett, Phys. Rev. B $\bf{100}$, 134502 (2019).
\bibitem{chakrabortty2023effect} S. Chakrabortty, R. Kumar and N. Mohapatra, Phys. Rev. B $\bf{107}$, 024503 (2023).
\bibitem{manchon2015new} A. Manchon, H. C. Koo, J. Nitta, S. M. Frolov and R. A. Duine , Nat. Mater. $\bf{14}$, 871 (2015).
\bibitem{caviglia2010tunable} A. D. Caviglia, M. Gabay, S. Gariglio, N. Reyren, C. Cancellieri and J.-M. Triscone, Phys. Rev. Lett. $\bf{104}$, 126803 (2010).
\bibitem{governale2002quantum} M. Governale, Phys. Rev. Lett. $\bf{89}$, 206802 (2002).
\bibitem{yu2023chirality} T. Yu, Z. Luo and G. E.W. Bauer , Phys. Rep. $\bf{1009}$, 1 (2023).
\bibitem{kim2013chirality} K. W. Kim, H. W. Lee, K. J. Lee and M. D. Stiles, Phys. Rev. Lett. $\bf{111}$, 216601 (2013).
\bibitem{grytsiuk2020topological} S. Grytsiuk, J. P. Hanke, M. Hoffmann, J. Bouaziz, O. Gomonay, G. Bihlmayer, S. Lounis, Y. Mokrousov and S. Blügel , Phys. Rev. Lett. $\bf{11}$, 511 (2020).
\bibitem{banerjee2014enhanced} S. Banerjee, J. Rowland, O. Erten, and M. Randeria, Phys. Rev. X $\bf{4}$, 031045 (2014).
\bibitem{kim2017spin} K. W. Kim, K. J. Lee, J. Sinova, H. W. Lee and M. D. Stiles, Phys. Rev. B $\bf{96}$, 104438 (2017).
\bibitem{brataas2014spin} A. Brataas and K. M. D. Hals, Nat. Nanotechnol. $\bf{9}$, 86 (2014).
\bibitem{zhu2019spin} L. Zhu, D.C. Ralph and R. A. Buhrman, Phys. Rev. Lett.  $\bf{112}$, 077201 (2019).
\bibitem{baek2018spin} S. h. C. Baek, V. P. Amin, Y. W. Oh, G. Go, S. J. Lee, G. H. Lee, K. J. Kim, M. D. Stiles, B. G. Park and K. J. Lee, Nat. Mater. $\bf{17}$, 509 (2018).
\bibitem{ciccarelli2016room} C. Ciccarelli, L. Anderson, V. Tshitoyan, A. J. Ferguson, F. Gerhard, C. Gould, L. W. Molenkamp, J. Gayles, J. Železný, L. Šmejkal, Z. Yuan, J. Sinova, F. Freimuth and T. Jungwirth , Nat. Phys. $\bf{12}$, 855 (2016).
\bibitem{tesavrova2013experimental} N. Tesařová, P. Němec, E. Rozkotová, J. Zemen, T. Janda, D. Butkovičová, F. Trojánek, K. Olejník, V. Novák, P. Malý and T. Jungwirth 
, Nat. Photonics $\bf{7}$, 492 (2013).
\bibitem{lee2021efficient} S. Lee, M. G. Kang, D. Go, D. Kim, J. H. Kang, T. Lee, G. H. Lee, J. Kang, N. J. Lee, Y. Mokrousov, S. Kim, K. J. Kim, K. J. Lee and B. G. Park
, Commun. Phys. $\bf{4}$, 234 (2021).
\bibitem{song2021spin} C. Song, R. Zhang, L. Liao, Y. Zhou, X. Zhou, R. Chen, Y. You, X. Chen and F. Pan, Prog. Mater. Sci. $\bf{118}$, 100761 (2021).
\bibitem{li2019manipulation} Y. Li, K. W. Edmonds, X. Liu, H. Zheng and K. Wang, Adv. Quantum Technol. $\bf{2}$, 1800052 (2019).
\bibitem{kane2005quantum} C. L. Kane and E. J. Mele, Phys. Rev. Lett. $\bf{95}$, 226801 (2005).
\bibitem{hoffmann2013spin} A. Hoffmann, IEEE Trans. Magn. $\bf{49}$, 5172 (2013).
\bibitem{nadj2010spin} S. N. Perge, S. M. Frolov, E. P. A. M. Bakkers and L. P. Kouwenhoven, Nature $\bf{468}$, 1084 (2010).
\bibitem{hasan2010colloquium} M. Z. Hasan and C. L. Kane, Rev. Mod. Phys. $\bf{82}$, 3045 (2010).
\bibitem{qi2011topological} X. L. Qi and S. C. Zhang, Rev. Mod. Phys. $\bf{83}$, 1057 (2011).
\bibitem{beugeling2012topological} W. Beugeling, N. Goldman and C. M. Smith, Phys. Rev. B $\bf{86}$, 075118 (2012).
\bibitem{zhang2022topological} X. Zhang, J. Liu and F. Liu, Nano Lett. $\bf{22}$, 9000 (2022).
\bibitem{kim2022three} M. Kim, Z. Wang, Y. Yang, H. T. Teo, J. Rho and B. Zhang, Nat. Commun. $\bf{13}$, 3499 (2022).
\bibitem{young2011theoretical} S. M. Young, S. Chowdhury, E. J. Walter, E. J. Mele, C. L. Kane and A. M. Rappe, Phys. Rev. B $\bf{84}$, 085106 (2011).
\bibitem{island2019spin} J. O. Island, X. Cui, C. Lewandowski, J. Y. Khoo, E. M. Spanton, H. Zhou, D. Rhodes, J. C. Hone, T. Taniguchi, K. Watanabe, L. S. Levitov, M. P. Zaletel and A. F. Young , Nature $\bf{571}$, 85 (2019).
\bibitem{kheirkhah2020first} M. Kheirkhah, Z. Yan, Y. Nagai and F. Marsiglio, Phys. Rev. Lett. $\bf{125}$, 017001 (2020).
\bibitem{fang2022ferromagnetic} S. Fang, L. Ye, M. P. Ghimire, M. Kang, J. Liu, M. Han, L. Fu, M. Richter, J. Brink, E. Kaxiras, R. Comin and J. G. Checkelsky, Phys. Rev. B $\bf{105}$, 035107 (2022).
\bibitem{rademaker2022spin} L. Rademaker, Phys. Rev. B $\bf{105}$, 195428 (2022).
\bibitem{Dejean2022} N. T. Dejean, F. G. Eich and A. Rubio , Npj Comput. Mater. $\bf{8}$, 145 (2022).
\bibitem{tian2020spin} L. Tian, Y. Liu, W. Meng, X. Zhang, X. Dai and G. Liu, J. Phys. Chem. Lett. $\bf{11}$, 10340 (2020).
\bibitem{pesin2012spintronics} D. Pesin and A. H. MacDonald, Nat. Mater. $\bf{11}$, 409 (2012).
\bibitem{ezawa2015monolayer}D. Pesin and A. H. MacDonald, J. Phys. Soc. Jpn. $\bf{84}$, 121003 (2015).
\bibitem{bian2016topological} G. Bian, T. R. Chang, R. Sankar, S. Y. Xu, H. Zheng, T. Neupert, C. K. Chiu, S. M. Huang, G. Chang, I. Belopolski, D. S. Sanchez, M. Neupane, N. Alidoust, C. Liu, B. Wang, C. C. Lee, H. T. Jeng, C. Zhang, Z. Yuan, S. Jia, A. Bansil, F. Chou, H. Lin and M. Z. Hasan,  Nat. Commun. $\bf{7}$, 10556 (2016).
\bibitem{singh2018spin} B. Singh, S. Mardanya, C. Su, H. Lin, A. Agarwal and A. Bansil, Phys. Rev. B $\bf{98}$, 085122 (2018).
\bibitem{howlader2020strong} S. Howlader, S. Saha, R. Kumar, V. Nagpal, S. Patnaik, T. Das and G. Sheet, Phys. Rev. B $\bf{102}$, 104434 (2020).
\bibitem{sheng2014topological} X. L. Sheng, Z. Wang, R. Yu, H. Weng, Z. Fang and X. Dai, Phys. Rev. B $\bf{90}$, 245308 (2014).

\bibitem{dziawa2012topological} P. Dziawa, B. J. Kowalski, K. Dybko, R. Buczko, A. Szczerbakow, M. Szot, E. Łusakowska, T. Balasubramanian, B. M. Wojek, M. H. Berntsen, O. Tjernberg and T. Story , Nat. Mater. $\bf{11}$, 1023 (2012).
\bibitem{mahfouzi2012spin} F. Mahfouzi, N. Nagaosa and B. K. Nikolić, Phys. Rev. Lett.  $\bf{109}$, 166602 (2012).
\bibitem{naimer2023twist} T. Naimer and J. Fabian, Phys. Rev. B $\bf{107}$, 195144 (2023).



\bibitem{belayadi2023spin} A Belayadi and P Vasilopoulos, Nat. Nanotechnol. $\bf{34}$, 365706 (2023).
\bibitem{zholudev2012magnetospectroscopy} M. Zholudev, F. Teppe, M. Orlita, C. Consejo, J. Torres, N. Dyakonova, M. Czapkiewicz, J. Wróbel, G. Grabecki, N. Mikhailov, S. Dvoretskii, A. Ikonnikov, K. Spirin, V. Aleshkin, V. Gavrilenko and W. Knap, Phys. Rev. B $\bf{86}$, 205420 (2012).
\bibitem{li2022electronic} Y. Li and G. Xu, Phys. Rev. Mater. $\bf{6}$, 104201 (2022).
\bibitem{de2020jacutingaite} F. C. D. Lima, R. H. Miwa and A. Fazzio, Phys. Rev. B $\bf{102}$, 235153 (2020).
\bibitem{liu2017two} P. F. Liu , L. Zhou , S. Tretiak  and L. M. Wu , J. Mater. Chem. C $\bf{5}$, 9181 (2017).
\bibitem{virot2013engineering} F. Virot, R. Hayn, M. Richter and J. Brink, Phys. Rev. Lett. $\bf{111}$, 146803 (2013).
\bibitem{mughal1969preparation} S. A. Mughal, A. J. Payne and B. Ray, J. Mater. Sci. $\bf{4}$, 895 (1969).
\bibitem{verozubova2010growth} G. A. Verozubova, A.O. Okunev, A.I. Gribenyukov, A.Y. Trofimiv, E.M. Trukhanov and A.V. Kolesnikov, J. Cryst. Growth $\bf{312}$, 1122 (2010).
\bibitem{masumoto1966preparation} K. Masumoto, S. Isomura and W. Goto, J. Phys. Chem. Solids $\bf{27}$, 1939 (1966).
\bibitem{rubenstein1968preparation} M. Rubenstein and R W  Ure Jr., J. Phys. Chem. Solids $\bf{29}$, 551 (1968).
\bibitem{zawilski2010growth} K. T. Zawilski, P. G. Schunemann, T. C. Pollak, D. E. Zelmon, N. C. Fernelius and F. K. Hopkins, J. Cryst. Growth $\bf{312}$, 1127 (2010).
\bibitem{zhang2012growth} G. Zhang, X. Tao, H. Ruan, S. Wang and Q. Shi, J. Cryst. Growth $\bf{340}$, 197 (2012).
\bibitem{vohl1979synthesis} P. Vohl, J. Electron. Mater. $\bf{8}$, 517 (1979).
\bibitem{buehler1971concerning} E. Buehler and J.H. Wernick, J. Cryst. Growth $\bf{8}$, 324 (1971).
\bibitem{shirakata1990growth} S. Shirakata and S. Isomura,  J. Cryst. Growth $\bf{99}$, 781 (1990).
\bibitem{trykozko1975ternary} R.T. Trykozko, Mater. Res. Bull. $\bf{10}$, 489 (1975).
\bibitem{khan2020review} I. S. Khan, K. N. Heinselman and A. Zakutayev, J. Phys. Energy $\bf{2}$, 032007 (2020).
\bibitem{kawamura2021synthesis} F. Kawamura, H. Murata, M. Imura, N. Yamada and T. Taniguchi, Inorg. Chem. $\bf{60}$, 1773 (2021).
\bibitem{greenaway2020combinatorial} A. L. Greenaway, A. L. Loutris, K. N. Heinselman, C. L. Melamed, R. R. Schnepf, M. B. Tellekamp, R. W. Robinson, R. Sherbondy, D. Bardgett, S. Bauers, A. Zakutayev, S. T. Christensen, S. Lany and A. C. Tamboli, J. Am. Chem. Soc. $\bf{142}$, 8421 (2020).
\bibitem{rom2021bulk} C. L. Rom, M. J. Fallon, A. Wustrow, A. L. Prieto and J. R. Neilson,Chem. Mater. $\bf{33}$, 5345 (2021).
\bibitem{esmaeilzadeh2006crystal} S. Esmaeilzadeh, U. Hålenius and M. Valldor, Chem. Mater. $\bf{18}$, 2713 (2006).


\bibitem{fan2014energetic} F. J.Fan, L. Wu and S. H. Yu ,  Energy Environ. Sci. $\bf{7}$, 190 (2014).
\bibitem{ohmer1998emergence} M. C. Ohmer and R. Pandey, MRS Bull. $\bf{23}$, 16 (1998).
\bibitem{sadhukhan2020first} B. Sadhukhan, Y. Zhang, R. Ray and J. Brink, Phys. Rev. Mater. $\bf{4}$, 064602 (2020).


\bibitem{momma2008vesta}\textcolor{black}{K. Momma and F. Izumi, J. Appl. Crystallogr.$\bf{41}$, 653 (2008).}


\bibitem{vasp1} G. Kresse and J. Hafner, Phys. Rev. B $\bf{47}$, 558(R) (1993).
\bibitem{vasp2} G. Kresse and J. Furthmuller, Phys. Rev. B $\bf{54}$, 11169 (1996).
\bibitem{PBE} J. P. Perdew, K. Burke, and M. Ernzerhof, Phys. Rev. Lett. $\bf{77}$, 3865 (1996).
\bibitem{wannier1} N. Marzari and D. Vanderbilt, Phys. Rev. B $\bf{56}$, 12847 (1997).
\bibitem{wannier2} I. Souza, N. Marzari, and D. Vanderbilt, Phys. Rev. B $\bf{65}$, 035109 (2001).
\bibitem{wannier3} G. Pizzi, V. Vitale, R. Arita, S. Blügel, F. Freimuth, G. Géranton, M. Gibertini, D. Gresch, C. Johnson, T. Koretsune, et al., J. of Phys.: Cond. Matt. $\bf{32}$, 165902 (2020).
\bibitem{wanniertool} Q. Wu, S. Zhang, H.-F. Song, M. Troyer, and A. A. Soluyanov, Computer Physics Communications $\bf{224}$, 405 (2018).
\bibitem{suh2004combinatorial} C. Suh and K. Rajan, Appl. Surf. Sci. $\bf{223}$, 148 (2004).

\bibitem{basalaev2020simulation} Y. M. Basalaev and M. Y. Basalaeva, J. Struct. Chem. $\bf{61}$, 1007 (2020).

\bibitem{zhong2019first} Y. Zhong, H. Mei, D. He, X. Du and N. Cheng, J. Phys. Chem. Solids $\bf{134}$, 157 (2019).

\bibitem{jaffe1984theory} J. E. Jaffe and A. Zunger,Phys. Rev. B  $\bf{29}$, 1882 (1984).
\bibitem{Surasree2022} S. Sadhukhan, B. Sadhukhan and S. Kanungo, Phys. Rev. B $\bf{106}$, 125112 (2022).



\bibitem{mouhat2014necessary} \textcolor{black}{F. Mouhat and F. X. Coudert, Phys. Rev. B $\bf{90}$, 224104 (2014).}


\bibitem{baghdad2022study} \textcolor{black}{A. H. Baghdad, H. Bouafia, B. Djebour, B. Sahli, S. Hiadsi, M. Elchikh and M. Attou , J. Phys. Chem. Solids $\bf{167}$, 110756 (2022).}

\bibitem{ruan2016ideal} J. Ruan, S. K. Jian, D. Zhang, H. Yao, H. Zhang, S. C. Zhang and D. Xing,Phys. Rev. Lett. $\bf{116}$, 226801 (2016).

\bibitem{kanungo2016weak} S. Kanungo, K. Mogare, B. Yan, M. Reehuis, A. Hoser, C. Felser and M. Jansen, Phys. Rev. B $\bf{93}$, 245148 (2016).
\bibitem{kanungo2022SOC} R. Roy and S. Kanungo, Phys. Rev. B $\bf{106}$, 125113 (2022).


\end{thebibliography}
\end{document}